\documentclass[journal]{IEEEtran}

\usepackage{myStyleIEEE}

\DeclareSIUnit \voltampere { VA } %apparent power 
\DeclareSIUnit \var { var } %volt-ampere reactive - idle power 

\IEEEoverridecommandlockouts

\begin{document}

\title{Effect of Dispatch Decisions on Small-Signal Stability of Converter-Dominated Power Systems}

\renewcommand{\theenumi}{\alph{enumi}}

\author{
\IEEEauthorblockN{Maitraya Avadhut Desai\IEEEauthorrefmark{1}, Ognjen Stanojev\IEEEauthorrefmark{2}, Simon~Muntwiler\IEEEauthorrefmark{3}, Gabriela Hug\IEEEauthorrefmark{1}}%

\IEEEauthorblockA{\IEEEauthorrefmark{1} EEH - Power Systems Laboratory, ETH Zurich, Switzerland} %

\IEEEauthorblockA{\IEEEauthorrefmark{2} ABB Corporate Research Center, Switzerland}\,\,\,
\IEEEauthorblockA{\IEEEauthorrefmark{3} ETH Sustainability, ETH Zurich, Switzerland} %

Emails: \{desai, hug\}@eeh.ee.ethz.ch, ognjen.stanojev@ch.abb.com, simonmu@ethz.ch
\thanks{Research supported by NCCR Automation, a National Centre of Competence in Research, funded by the Swiss National Science Foundation (grant number 51NF40\_180545).}
}
% \author{Maitraya~Avadhut~Desai,~\IEEEmembership{Graduate~Student~Member,~IEEE,}
%         Ognjen~Stanojev,~\IEEEmembership{Graduate~Student~Member,~IEEE,}
%         Simon~Muntwiler,~\IEEEmembership{Graduate~Student~Member,~IEEE,}
%         Gabriela~Hug,~\IEEEmembership{Senior~Member,~IEEE}% <-this % stops a space

% \thanks{This research was supported by the Swiss National Science Foundation under NCCR Automation, grant agreement 51NF40\_180545.}
% \thanks{M. A. Desai, O. Stanojev and G. Hug are with the Power Systems Laboratory at ETH Z\"{u}rich, Z\"{u}rich, 8092,  Switzerland, emails:\{desai, stanojev,hug\}@eeh.ee.ethz.ch. \\
% S. Muntwiler is with the Institute for Dynamic Systems and Control ETH Z\"{u}rich, Z\"{u}rich, 8092,  Switzerland, email: simonmu@ethz.ch.\\ \textit{(Corresponding Author: M.A. Desai)}
% }}

\maketitle
\IEEEpeerreviewmaketitle

%captions, result first paragraph

%ABSTRACT
\begin{abstract}
Small-signal stability of modern converter-dominated power systems has been the subject of extensive research, particularly from the perspective of device-level control design for grid-forming (GFM) and grid-following (GFL) converters. However, the influence of power flow variables on system stability has received limited attention. Conventional small-signal stability analyses are typically conducted at a specific operating point, emphasizing the selection of control or system design parameters while neglecting the sensitivity of stability characteristics to operating conditions. This paper seeks to bridge this gap by systematically investigating the impact of dispatch decisions on the small-signal stability of converter-based power systems. Our findings are first illustrated on a three-bus system and then validated on the standard IEEE 39-bus test system to demonstrate scalability. Across the test systems, we find that high-voltage capacitive operation of GFL converters limits its active power injection, whereas inductive operation permits higher injections, and it is generally preferable for the GFM converter to supply more active power.
\end{abstract}

%INDEX TERMS

\section{Introduction} \label{sec:intro}
The global trend towards sustainable energy production has led to increased investments in converter-based generation. Consequently, the deployment of voltage source converters (VSCs) has increased substantially \cite{Milano2018}. Control strategies for VSCs are generally categorized into two classes: (i) grid-forming (GFM), which enforces stiff regulation of the interface voltage vector, and (ii) grid-following (GFL), which presumes a fixed interface voltage and aligns its angle using a synchronization mechanism \cite{Yitong_Duality_2022}. These paradigms shape the dynamic behavior of modern converter-dominated systems, leading to interaction mechanisms that differ markedly from those of conventional synchronous generator-based systems \cite{LowInertia_Dorfler_2023}.

Small-signal stability analysis is concerned with the analysis of the impact of small perturbations on system stability around an operating point and uncovering the underlying properties of the system at the operating point \cite{Kundur-book}.  Considerable attention has been devoted to the small-signal stability of converter-dominated power systems thus far \cite{Markovic_UnderstandingSSS, BenedettI_SSS_2021Powertech, Yang_SSS_2023, QuantInteract_Indla_TPS, Henriquez_Line_2020, GLDOC_PESGM_Kravis}. For example, \cite{Markovic_UnderstandingSSS} investigates stability primarily as a function of the penetration levels of GFM and GFL VSCs, while \cite{BenedettI_SSS_2021Powertech, Yang_SSS_2023} examine the sensitivity of stability to converter control gains. The interaction between VSC and synchronous generator modes is quantified in \cite{QuantInteract_Indla_TPS} through eigenvalue analysis, and the role of detailed component models is shown in \cite{Henriquez_Line_2020, GLDOC_PESGM_Kravis}. 

Nevertheless, all the above-mentioned studies focus on either determining the maximum permissible converter penetration in a system or parameter tuning at either the device- or system-level. The effect of the selected operating point (driven by converter setpoints and the underlying power flow variables) on small-signal stability has not been systematically and thoroughly investigated thus far. 
Preliminary investigations presented in \cite{Markovic_UnderstandingSSS} revealed that a synchronization-related mode tends to shift toward the right-half plane as the active power setpoint of a GFM converter increases in a two-bus system. This line of research was further explored in \cite{GLDOC_PESGM_Kravis}, where it was demonstrated that operating scenarios in a 9-bus system become unstable when GFL converters are assigned higher power setpoints than their GFM counterparts. On the other hand, \cite{Yunda2025} developed an analytical framework to characterize the stable operating region of a GFM converter connected to an infinite bus. This framework accounts for a wide range of power transfer angles and thus considers various aspects of the operating point influence.

While the aforementioned studies have provided valuable insights, several critical aspects remain insufficiently addressed. In particular, although a converter’s operating point is determined by its voltage, active power, and reactive power setpoints, prior analyses have primarily focused on the influence of power setpoints. To accurately characterize the dispatch limitations of VSCs, it is necessary to construct stable operating regions that incorporate constraints on all relevant power flow variables. Additionally, the influence of control and system parameters on small-signal stability is operating-point dependent, yet this dependence is often neglected.

In this paper, we extend the scope of previous studies presented in \cite{Markovic_UnderstandingSSS, BenedettI_SSS_2021Powertech, Yang_SSS_2023, QuantInteract_Indla_TPS, Henriquez_Line_2020, GLDOC_PESGM_Kravis,Yunda2025} by analyzing the effect of dispatch decisions on the small-signal stability of low-inertia systems, rather than focusing solely on power setpoints or specific control or system parameters. Furthermore, to enable a comprehensive characterization of stability-related dispatch constraints for GFM and GFL converters, we construct stable operating regions that incorporate limitations across all relevant power flow variables. Finally, we demonstrate how the stable operating ranges change if different sets of control parameters are used, and thus demonstrate the impact of controller tuning, connecting our findings to the conclusions established in the literature. Our findings are first showcased on a three-bus system and then validated on the standard IEEE 39-bus test system to demonstrate their scalability and practical relevance.

%The remainder of this paper is structured as follows. Section~\ref{sec:model} briefly describes the considered dynamic model of the converter-dominated power system. The effect of the power flow variables on small-signal stability is presented in Section~\ref{sec:Effect}. Finally, Section~\ref{sec:conclusion} provides the conclusions of the paper.

\section{Modeling of Converter-Dominated Systems} \label{sec:model}
This section presents an overview of the differential algebraic equation (DAE) models used to describe typical components in converter-dominated power systems. The dynamic behavior of transmission lines and detailed representations of generating units, including synchronous generators, GFM and GFL converters, are considered. All models are formulated in the synchronously rotating reference frame (SRF) and expressed in the per-unit system to facilitate analysis and comparison.

\subsection{Graph-theoretic Network Modeling \& Line Dynamics}

We consider a transmission power system represented by a connected graph, denoted by $\mathcal{G}(\mathcal{N},\mathcal{E})$, where $\mathcal{N}$ represents the set of network nodes and $\mathcal{E} \subseteq \mathcal{N}\times\mathcal{N}$ denotes the set of network edges. The set of nodes is partitioned as $\mathcal{N} = \mathcal{N}_\mathrm{SG}\, \cup\, \mathcal{N}_\mathrm{GFM} \,\cup\, \mathcal{N}_\mathrm{GFL} \,\cup\, \mathcal{N}_\mathrm{Load}$ to define sets of nodes that host synchronous generators, GFM converters, GFL converters and loads, respectively. For every node $k\in\mathcal{N}$ in the graph, let $\vect{v}_{{\rm{n}}_k}\in\R^2$ denote the associated voltage vector (consisting of a $v_{{\rm{n}}_k}^{\rm{d}}$ and a $v_{{\rm{n}}_k}^{\rm{q}}$ component). For each line $(k,l)\in\mathcal{E}$, let $r_{kl}\in\R_{\geq 0}$ and $\ell_{kl}\in\R_{\geq 0}$ represent its respective resistance and inductance values, and $\vect{i}_{kl}\in\R^2$ represent the corresponding branch current vector. The transmission lines of the network are modeled as $\pi$-sections in a $dq$-frame rotating at frequency $\omega_{\rm g}$. The differential equations for the branch current $\vect{i}_{kl}\,\forall \, (k,l)\in\mathcal{E}$ are represented as
\begin{equation}
    \ddt\vect{i}_{kl}=\frac{\omega_{\rm b}}{\ell_{kl}}(\vect{v}_{{\rm{n}}_k}-\vect{v}_{{\rm{n}}_l})-\left(\frac{r_{kl}}{\ell_{kl}}\omega_{\rm b}+\mat{\mathcal{J}}\omega_{\rm b} \omega_{\rm g}\right)\vect{i}_{kl},\label{eq:lineCurrent}
\end{equation}
where $\omega_{\rm b}$ represents the base frequency. The $2\times 2$, rotation matrix at an angle $\theta$ is denoted as $\mat{{\mathcal{R(\theta)}}}$ and $\mat{\mathcal{J}}=\mat{{\mathcal{R}}}\,(\pi/2)$ can be interpreted as the embedding of the complex unit $j=\sqrt{-1}$ in $\mathbb{R}^2$. The dynamic equation of the voltage $\vect{v}_{{\rm{n}}_k},\,\forall\, k\in\mathcal{N}$ is given as 
\begin{equation}
    \ddt\vect{v}_{{\rm{n}}_k}=\frac{\omega_{\rm b}}{c_{k}}\vect{i}_{{\rm{c}}_k}-\left(\frac{g_{k}}{c_{k}}\omega_{\rm b}+\mat{\mathcal{J}}\omega_{\rm b} \omega_{\rm g}\right)\vect{v}_{{\rm{n}}_k},\label{eq:nodeVoltage}
\end{equation}
where $c_k$, $g_k$ and $\vect{i}_{{\rm{c}}_k}$ represent the aggregated shunt capacitance, aggregated shunt conductance and the current flowing through the shunt capacitance at node $k$. A constant impedance model is used for all loads.

\subsection{Power Converter Models} \label{subsec:sys_level}
\begin{figure}[!t]
    \centering
    \includegraphics[scale=0.9]{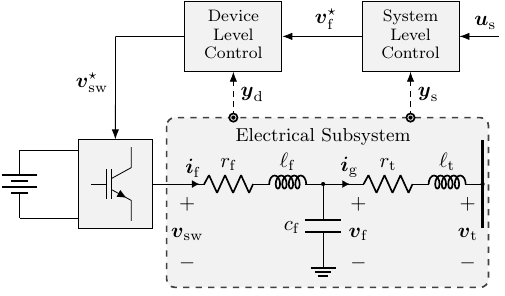}
    \caption{General converter configuration scheme.}
    \label{fig:converter_diag}
    \vspace{-0.6cm}
\end{figure}

The converter model under consideration is depicted in Fig.~\ref{fig:converter_diag}. It comprises a two-level cascaded control architecture, a switching stage supplied by a constant DC input voltage, and an AC subsystem incorporating an RLC filter $(r_{\rm f},\ell_{\rm f},c_{\rm f})$ with an equivalent transformer model $(r_{\rm t},\ell_{\rm t})$. Within this control framework, the outer \textit{system-level} control layer generates a reference signal $\vect{v}_{\rm f}^\star$ for the converter’s output voltage, which is subsequently regulated by the \textit{device-level} controller by adapting the switching voltage $\vect{v}_\mathrm{sw}= \vect{v}_\mathrm{sw}^\star$.

\subsubsection{System-level Control} The input measurement vector of the system-level controller is defined as $\vect{y}_{\rm s}=(\vect{v}_{\rm f},\vect{i}_{\rm g})\in\R^4$, where $\vect{v}_{\rm f} \in \R^2$ denotes the filter voltage and $\vect{i}_{\rm g} \in \R^2$ denotes the converter current injection into the system. We calculate the instantaneous active power $p_{\rm c}= \vect{v}^\mathsf{T}_{\rm f} \vect{i}_{\rm g}$ and reactive power $q_{\rm c} = \vect{v}^\mathsf{T}_{\rm f} \mat{\mathcal{J}}^\mathsf{T} \vect{i}_{\rm g}$ power using these measurements. Droop control is used on the active and reactive power imbalance with respect to the setpoints $(p_{\rm c}^\star,q_{\rm c}^\star)$ along with a low-pass filter, i.e., 
\begin{subequations} \label{eq:powersync1}
\begin{align} 
	 \ddt{\tilde{\omega}}_{\rm c} &= -\omega_{\rm z}\tilde{\omega}_{\rm c} + R_{\rm c}^{\rm p}\omega_{\rm z}(p_{\rm c}^\star-p_{\rm c}),\\
	 \ddt{\tilde{v}}_{\rm c}&= -\omega_{\rm z}\tilde{v}_{\rm c} + R_{\rm c}^{\rm q}\omega_{\rm z}(q_{\rm c}^\star-q_{\rm c}),
\end{align}
\end{subequations}
where $R_{\rm c}^{\rm p}\in\R_{>0}$, $R_{\rm c}^{\rm q}\in\R_{\geq 0}$ are the droop gains and $\omega_{\rm z}\in\R_{>0}$ denotes the low-pass filter cut-off frequency. The voltage reference $\vect{v}_{\rm c} \in \R^2$ in the $dq$-coordinates defined by $(\theta_{\rm c},\omega_{\rm c})$ can thereby be generated as
\begin{subequations} \label{eq:powersync2}
\begin{align} 
    \ddt{\theta}_{\rm c} &= \omega_{\rm b}\omega_{\rm c}, &\omega_{\rm c} = \omega_{\rm s} + \tilde{\omega}_{\rm c}, \\
    v_{\rm c}^{\rm q} &=0, &v_{\rm c}^{\rm d} = v_{\rm c}^\star + \tilde{v}_{\rm c},
\end{align}
\end{subequations}
where $\omega_{\rm s}$ represents the synchronization frequency and $v_{\rm c}^\star$ is the setpoint for the voltage magnitude. As an additional degree of freedom for stabilization and disturbance rejection, a virtual impedance $(r_{\rm v},\ell_{\rm v})\in\R^2_{\geq0}$ of the following form $\vect{v}_{\rm f}^\star = \vect{v}_{\rm c} - r_{\rm v} \vect{i}_{\rm g}  - \mat{\mathcal{J}}\omega_{\rm c} \ell_{\rm v}  \vect{i}_{\rm g}$ is commonly implemented.

Within the system-level control framework, a distinction is made between the GFL and GFM modes of operation.

\textit{Grid-Following Mode:} A key component of GFL converters is a synchronization device in the form of a phase locked loop (PLL), which estimates the phase angle $\theta_{\rm s} \in [-\pi,\pi)$ of the voltage $\vect{v}_{\rm f}$ and the frequency $\omega_{\rm s}\in\R_{>0}$:
\begin{equation}\label{eq:theta_pll}
    \omega_{\rm s} = \omega_{\rm g} + K_{\rm P}^{\rm s} v_{\rm f}^{\rm q} + K_{\rm I}^{\rm s} \varepsilon,\quad \ddt{\varepsilon} = v_{\rm f}^{\rm q},\quad
    \ddt{\theta}_{\rm s} = \omega_{\rm b}\omega_{\rm s},
\end{equation}
where $K_{\rm P}^{\rm s}\in\R_{>0}$, $K_{\rm I}^{\rm s}\in\R_{\geq 0}$ are the proportional and integral control gains of the synchronization unit, and $\varepsilon\in\R$ is the integrator state. Therefore, for GFL converters, the synchronization frequency $\omega_{\rm s}$ in \eqref{eq:powersync2} is defined by the PLL.

\textit{Grid-Forming Mode:} Unlike the GFL mode of operation, the GFM control does not require a synchronization unit, since $\omega_{\rm s}=\omega_{\rm c}^\star$ is assigned a constant reference value. Consequently, such units achieve self-synchronization with the power grid by adjusting the converter frequency and voltage in response to output power deviations, thereby eliminating the need for a PLL.

\subsubsection{Device-level Control}
Assuming a given voltage reference $\vect{v}^\star_{\rm f} \in \R^2$ in $dq$-coordinates defined by $(\theta_{\rm c},\omega_{\rm c})$, the device-level control is constructed in a dual-loop fashion, as a cascade of voltage and current controllers computing a switching voltage reference $\vect{v}_\mathrm{sw}^\star\in\R^2$:
\hspace{-5em}
\begin{subequations} \label{eq:srf_vi}
\begin{align} 
    \ddt{\vect{\xi}} &=  \vect{v}^\star_{\rm f} - \vect{v}_{\rm f}, \\
    \vect{i}_{\rm f}^\star &= K_{\rm P}^{\rm v} (\vect{v}^\star_{\rm f} - \vect{v}_{\rm f}) + K_{\rm I}^{\rm v} \vect{\xi} + K_{\rm F}^{\rm v} \vect{i}_{\rm g} + \mat{\mathcal{J}}\omega_{\rm c} c_{\rm f} \vect{v}_{\rm f}, \label{eq:srf_v_b}
    \end{align}
providing an internal current reference $i_f^\star$ followed by
\begin{align} 
    \ddt{\vect{\gamma}} &= \vect{i}_{\rm f}^\star - \vect{i}_{\rm f},\\
    \vect{v}_\mathrm{sw}^\star &= K_{\rm P}^{\rm i} (\vect{i}_{\rm f}^\star - \vect{i}_{\rm f}) + K_{\rm I}^{\rm i} \vect{\gamma} + K_{\rm F}^{\rm i} \vect{v}_{\rm f} + \mat{\mathcal{J}}\omega_{\rm c} \ell_{\rm f} \vect{i}_{\rm f}, \label{eq:srf_i_b}
    \end{align}
\end{subequations}
where $(K_{\rm P}^{\rm v},K_{\rm P}^{\rm i})\in\R^2_{>0}$, $(K_{\rm I}^{\rm v},K_{\rm I}^{\rm i})\in\R^2_{\geq0}$ and $(K_{\rm F}^{\rm v},K_{\rm F}^{\rm i})\in\mathbb{Z}_{\{0,1\}}^2$ are the respective proportional, integral, and feed-forward gains, $\vect{\xi}\in\R^2$ and $\vect{\gamma}\in\R^2$ represent the integrator states, and the superscripts $\rm v$ and $\rm i$ indicate the voltage and current controllers, respectively.

\subsection{Synchronous Generator Model}
We consider the detailed $8^{\rm{th}}$-order Sauer and Pai model for a synchronous generator that includes the round rotor model with its circuit dynamics, motion dynamics, and also considers stator dynamics. In addition, a prime mover and a governor of type TGOV1 along with an IEEE DC1A automatic voltage regulator (AVR) are also modeled. Similarly to the converters, the synchronous generator is interfaced to the grid through a transformer and modeled in an SRF. We refer the reader to reference \cite{Kundur-book} for a deeper understanding of the detailed model.

\subsection{Complete Model}
To obtain a consistent model of the entire system, it is imperative to perform a rotational transformation on the terminal quantity $\vect{x}_{{\rm t}_k}\,\in\,\{\vect{v}_{{\rm t}_k},\,\vect{i}_{{\rm g}_k}\}$ to obtain the respective network nodal quantity $\vect{x}_{{\rm n}_k}\,\in\,\{\vect{v}_{{\rm n}_k},\,\vect{i}_{{\rm n}_k}\}$ for all nodes $k\in\mathcal{N}$. This is done as follows:
\begin{equation}
    \vect{x}_{{\rm n}_k}=\mat{\mathcal{R}}(\theta_{k}-\theta_{\rm g})\vect{x}_{{\rm t}_k},\quad \forall \, k\in\mathcal{N} ,\label{eq:alignment}
\end{equation}
where the network's SRF speed is represented as $\ddt{\theta_{\rm g}}=\omega_{\rm b}\omega_{\rm g}$ and is usually chosen to be equal to that of an arbitrary converter or synchronous generator. The speed of the node $\ddt{\theta_{k}}=\omega_{k}$ is the SRF speed of the respective unit at the node. It is important to note here that the power flow variables are algebraic variables at the nodal level, i.e., 
\begin{subequations} \label{eq:powerflow}
\begin{align} 
p_k&=\vect{v}_{{\rm n}_k}^{\mathsf{T}}\vect{i}_{{\rm n}_k}, &q_k=\vect{v}_{{\rm n}_k}^{\mathsf{T}}{\mat{\mathcal{J}}}^{\mathsf{T}}\vect{i}_{{\rm n}_k}, \\
    v_{{\rm n}_k}&=\norm{\vect{v}_{{\rm n}_k}}, &\delta_{{\rm n}_k}=\tan^{-1}(v_{{\rm n}_k}^{\rm q}/v_{{\rm n}_k}^{\rm d}),
\end{align}
\end{subequations}
where the power injection at node $k$ is the generated power subtracted by the power consumed by the load at that node.

Finally, collecting the differential and algebraic variables of the generating units and the network in the vectors $\vect{x}$ and $\vect{z}$, we obtain a DAE system of the form
\begin{equation} 
    \ddt{\vect{x}}=\vect{f}\bigl(\vect{x},\,\vect{z}\bigr), \quad 0=\vect{g}\bigl(\vect{x},\,\vect{z}\bigr)\label{eq:DAEsys}.
\end{equation}

\subsection{Small-Signal Model}
To obtain the small-signal model of the power system, we linearize \eqref{eq:DAEsys} around a desired equilibrium $\vect{\rho}_{\rm o}=(\vect{x}_{\rm o},\vect{z}_{\rm o})$ such that $\vect{f}\bigl(\vect{\rho}_{\rm o}\bigr)=0$ and $\vect{g}\bigl(\vect{\rho}_{\rm o}\bigr)=0$. This is done by performing a power flow computation and using the resulting system state as the operating point. The result is a linearized DAE of the form
\begin{equation} 
    \ddt{\vect{\delta{x}}}=\mat{A}_{\rm xx}\vect{\delta{x}}+\mat{A}_{\rm xz}\vect{\delta{z}},\quad
    0=\mat{A}_{\rm zx}\vect{\delta{x}}+\mat{A}_{\rm zz}\vect{\delta{z}}, \label{eq:linDAEsys}
\end{equation}
where the respective deviations from the desired operating point are represented by ${\vect{\delta{x}}}=\vect{x}-\vect{x}_{\rm o},\,{\vect{\delta{z}}}=\vect{z}-\vect{z}_{\rm o}$. The matrices $\mat{A}_{\rm xx}$ and $\mat{A}_{\rm xz}$ denote the Jacobians of $f$ with respect to $\vect{x}$ and $\vect{z}$, respectively, and the matrices $\mat{A}_{\rm zx}$ and $\mat{A}_{\rm zz}$ denote the Jacobians of $g$ with respect to $\vect{x}$ and $\vect{z}$, respectively. The Jacobians are evaluated at $\vect{\rho}_{\rm o}$. Considering that the DAE \eqref{eq:DAEsys} is index-1, it can be represented by a linear ordinary differential equation (ODE) of the form 
\begin{equation} \label{eq:ODEsys}
    \ddt{\vect{\delta{x}}}=\underbrace{\bigl(\mat{A}_{\rm xx}-\mat{A}_{\rm xz}\mat{A}_{\rm zz}^{-1}\mat{A}_{\rm zx}\bigr)}_{\mat{\tilde{A}}}\vect{\delta{x}},
\end{equation}
where ${\mat{\tilde{A}}}\in\mathbb{R}^{n_x\times{n_x}}$ represents the reduced state-space matrix and $n_x$ denotes the total number of differential states. To determine whether the system is small-signal stable, the eigenvalues of the reduced state-space matrix, i.e.,  ${\vect{\lambda}(\mat{\tilde{A}})}$ are analyzed. If all eigenvalues have negative real parts, the power system is considered small-signal stable at the equilibrium $\vect{\rho}_{\rm o}$. 

\section{Results}\label{sec:Effect}

All described models are implemented in Python using Casadi's symbolic framework \cite{Andersson2019}. Time domain simulations are performed employing the collocation method with a time step of 0.0001$\,$s \footnote{The codebase will be made public upon acceptance}.
\begin{figure}[t]
    \centering
    \includegraphics[scale=0.7]{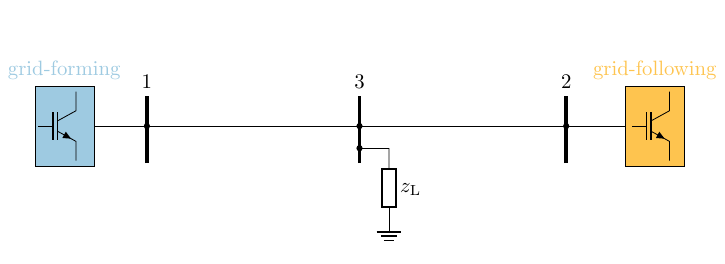}
    \caption{Three-bus system with a GFM and GFL converter at Bus 1 and Bus 2, respectively.}
    \label{fig:3bus_diag_1}
    \vspace{-0.5cm}
\end{figure}

\subsection{Three-bus system}

To highlight the effect of power flow variables on small-signal stability, we consider a simple case of a three-bus system shown in Fig.~\ref{fig:3bus_diag_1} $(\mathcal{N}=\{1,2,3\},\,\mathcal{E}=\{(1,3),(2,3)\})$. A GFM converter at Bus~1 and a GFL converter at Bus~2 feed a load at Bus~3 that is set to consume $1\,\rm{pu}$ active power and $0.1\,\rm{pu}$ reactive power. The parameters of the GFL converter and the GFM converter are adopted from \cite[Table~1, Table~2]{Markovic_UnderstandingSSS}. The line parameters are $r_{kl}=0.0146\,\rm{pu}$, $\ell_{kl}=0.146\,\rm{pu}$, $g_{kl}=0.05\,\rm{pu}$ and $c_{kl}=0.09\,\rm{pu}$ $\forall\, (k,l)\in\mathcal{E}$.

% To highlight the effect of the power flow variables on small-signal stability, we present case-studies using the three-bus system shown in Fig.~\ref{fig:3bus_diag_1} $(\mathcal{N}=\{1,2,3\},\,\mathcal{E}=\{(1,3),(2,3)\})$. The parameters of the GFL converter, the GFM converter, and the synchronous generator are adopted from \cite[Table~1, Table~2]{Markovic_UnderstandingSSS}. The load is set to consume $1\,\rm{pu}$ active power at nominal voltage. The line parameters are $r_{kl}=0.0293\,\rm{pu}$ and $\ell_{kl}=0.293\,\rm{pu},\,\forall\, (k,l)\in\mathcal{E}$. While we acknowledge the impact of device- and system-level parameter tuning on small-signal stability, we seek to motivate that the effect of the dispatch decision is commensurate. 

% In particular, we are interested in the generator dispatch and the effect that it has on the small-signal stability of the system.   It is important to acknowledge the effect that the controller gains or the system parameters have on small-signal stability. However, we seek to highlight that the effect of dispatch is commensurate.   

\subsubsection{PV Dispatch of Grid-Following Converter}

\begin{figure}[t]
    \centering
    \includegraphics[scale=0.99]{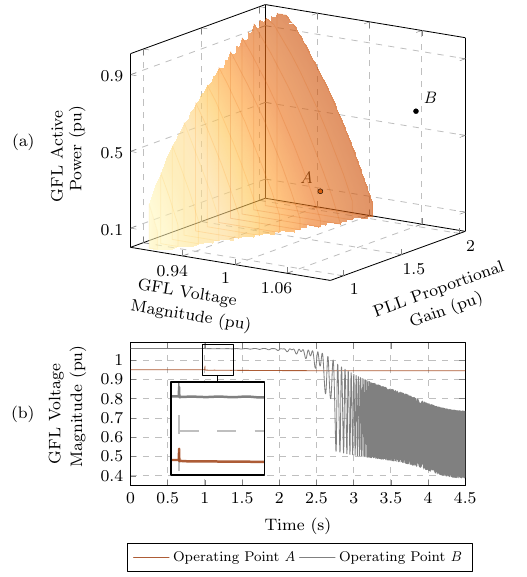}
    \caption{(a) Shaded region of stability with respect to the GFL active power injection and voltage magnitude as the PLL proportional gain is varied in the 3-bus system. (b) Time domain simulation for operating points a and b for a load disturbance at $t=1.00\,\rm{s}$.}
    \label{fig:3bus_PV_result}
    \vspace{-0.5cm}
\end{figure}

First, we analyze the small-signal stability of the system for the active power and voltage magnitude of the GFL converter in the range $p_2\in[0.01\,\rm{pu},1.00\,\rm{pu}]$ and $v_2\in[0.90\,\rm{pu},\,1.10\,\rm{pu}]$. Moreover, we also study the impact of a variation in the proportional gain of the PLL $K_{\rm P}^{\rm s}\in[0.90\,\rm{pu},\,2.00\,\rm{pu}]$. The eigenvalues of the system are evaluated to check for stability and the results of the analysis are shown in Fig.~\ref{fig:3bus_PV_result}. 

The shaded stability region in Fig.~\ref{fig:3bus_PV_result}(a) illustrates that small-signal stability depends on the operating point. This highlights the limitation of small-signal analyses that consider only controller parameters and disregard dispatch decisions. Two broad trends emerge. First, for any fixed proportional gain of the PLL, the admissible active power generation increases as the terminal voltage magnitude drops below the nominal value of $v_2=1.00\,\rm{pu}$ and decreases when the terminal voltage magnitude rises above the nominal value. Hence, to preserve stability, the GFL converter should be derated during over-voltage conditions. Second, increasing the proportional gain of the PLL enlarges the stable region, allowing higher active power injections and a wider voltage range. We consider a stable and an unstable operating point, Points \emph{A} and \emph{B}, respectively, as shown in Fig.~\ref{fig:3bus_PV_result}. At Point~\emph{A}, the GFL converter injects $p_2=0.10\,\rm{pu}$ at $v_2=0.95\,\rm{pu}$ and at Point~\emph{B} this converter injects $p_2=0.60\,\rm{pu}$ at $v_2=1.06\,\rm{pu}$. The PLL proportional gain is set as $K_{\rm P}^{\rm s}=2.00$ for each operating point. We perform a time domain simulation for both operating points considering an active power disturbance in the load Bus~3 of $0.05\,\rm{pu}$ at $t=1.00\,\rm{s}$. As seen in Fig.~\ref{fig:3bus_PV_result}(b), this disturbance results in a stable response for Point~\emph{A}, however, oscillatory instability is observed for Point~\emph{B} after the small-signal disturbance. The oscillatory mode of instability outside the stable region is caused by the interaction of the synchronization loops of the GFM and the GFL converters. In particular, the PLL dynamics show a very high participation in this case.

\subsubsection{PQ Dispatch of Grid-Following Converter}

\begin{figure}[t]
    \centering
    \includegraphics[scale=0.99]{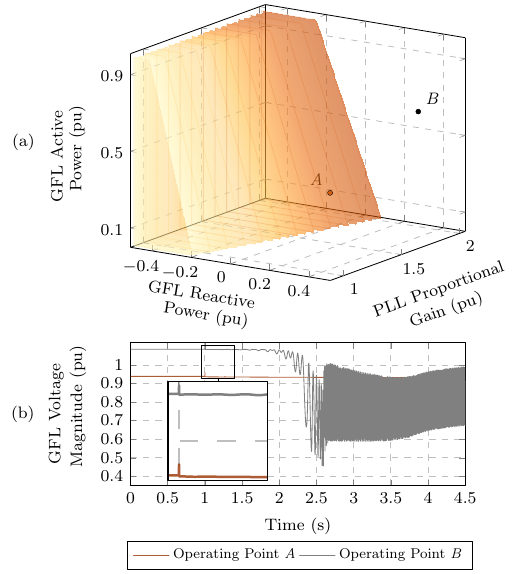}
    \caption{(a) Shaded region of stability with respect to the GFL active power and reactive power injection as the PLL proportional gain is varied in the 3-bus system. (b) Time domain simulation for operating points a and b for a load disturbance at $t=1.00\,\rm{s}$.}
    \label{fig:3bus_PQ_result}
    \vspace{-0.5cm}
\end{figure}

% \begin{figure}[!t]
% 	\centering
% 	\begin{minipage}{0.45\textwidth}
% 		\centering
%  		\hspace{-0.65cm}
% 		\scalebox{1.125}{\includegraphics[]{images/SectionIII/SG_GFM_3d_res_py.pdf}}\\
% 		\vspace{-0.1cm}  
% 	\end{minipage}
% 	\begin{minipage}{0.45\textwidth}
% 		\centering
%  		\hspace{-0.65cm}
% 		\scalebox{1.125}{\includegraphics[]{images/SectionIII/SG_GFM_2d_res_py.pdf}}\\
% 	\end{minipage}
% 	\caption{\label{fig:3bus_SG_GFM_res}Regions of stability with respect to active power and voltage magnitude as the length of the transmission line $2-3$ varies (top) and the region of stability projected onto the ($v_2,\,p_2$) plane (bottom) for the GFM converter.}
% 	\vspace{-0.35cm}
%\end{figure}

In the second analysis, the small-signal stability of the system is examined while varying the active power and the reactive power of the GFL converter in the range $p_2\in[0.01\,\rm{pu},1.00\,\rm{pu}]$ and $q_2\in[-0.50\,\rm{pu},\,0.50\,\rm{pu}]$. The proportional gain of the PLL is varied as $K_{\rm P}^{\rm s}\in[0.90\,\rm{pu},\,2.00\,\rm{pu}]$, similar to the previous analysis. The results of the analysis are shown in Fig.~\ref{fig:3bus_PQ_result}.

The shaded stability region in Fig.~\ref{fig:3bus_PQ_result}(a) shows that, for a fixed proportional PLL gain, the admissible active power injection grows as the bus absorbs reactive power and contracts as the bus supplies reactive power. The GFL converter can therefore inject more active power when operated with inductive reactive power dispatch, whereas capacitive dispatch necessitates derating of the converter unit. Similarly to the previous analysis, increasing the proportional gain of the PLL increases the size of the stable region, and therefore admits higher active power and reactive power dispatch of the converter.

To further analyze system behavior, we consider a stable operating Point~\emph{A}, where the GFL converter is set to dispatch $p_2=0.10\,\rm{pu}$ and $q_2=-0.15\,\rm{pu}$ and an unstable operating Point~\emph{B}, where the GFL converter is set to dispatch $p_2=0.60\,\rm{pu}$ and $q_2=0.30\,\rm{pu}$. For both operating points, the proportional PLL gain is set to $K_{\rm P}^{\rm s}=2.00$. The time-domain simulation for both operating points is shown in Fig.~\ref{fig:3bus_PQ_result}(b) where we consider an active power disturbance of $0.05\,\rm{pu}$ in the load at Bus~3 at $t=1.00\,\rm{s}$. While Point~\emph{A} results in a stable response, the oscillatory instability observed beyond the stable region is illustrated in the response of operating Point~\emph{B}.

In \cite{QuantInteract_Indla_TPS}, it has been reported that the interactions of the critical modes are higher for a lower bandwidth of the PLL. The examples we present extend this observation by showing how these interactions manifest for different dispatch decisions.

% The results of the analysis are shown in Fig.~\ref{fig:3bus_SG_GFM_res}. The stable operating region of the three-bus system contracts as the 2–3 line becomes longer. The contraction occurs primarily through a tighter upper bound on the active power dispatch of the GFM converter. Across the examined range, higher voltage magnitude offers a small increase in the admissible active power. At lower and near-nominal voltage magnitudes, a lower power dispatch of the GFM unit is clearly more favorable for stability, and this preference becomes increasingly pronounced as the line length increases. In \cite{BenedettI_SSS_2021Powertech}, it has been shown that lengthening the line can move modes unfavorably. The present example complements this insight by showing how the stability margin depends on the converter’s dispatch decision and how conservative active-power setpoints are especially beneficial on longer lines.

\subsection{IEEE 39-bus System}
To examine how our observations extend to larger and more realistic systems, we consider the IEEE 39-bus test system for which the original system parameters are given in \cite{benchmark_hiskens} and the synchronous generator parameters are adopted from \cite{milos}. Furthermore, we replace the synchronous generators at buses 31, 33, and 36 with GFM converters of nominal capacities $100\,\rm{MVA}$, $100\,\rm{MVA}$ and $1000\,\rm{MVA}$, respectively, and the synchronous generators at bus 30 with a GFL converter of nominal capacity $1000\,\rm{MVA}$. This ensures equitable distribution of conventional synchronous generator units and converter-interfaced units in the system.  

\begin{figure}[t]
    \centering
    \includegraphics[scale=0.8]{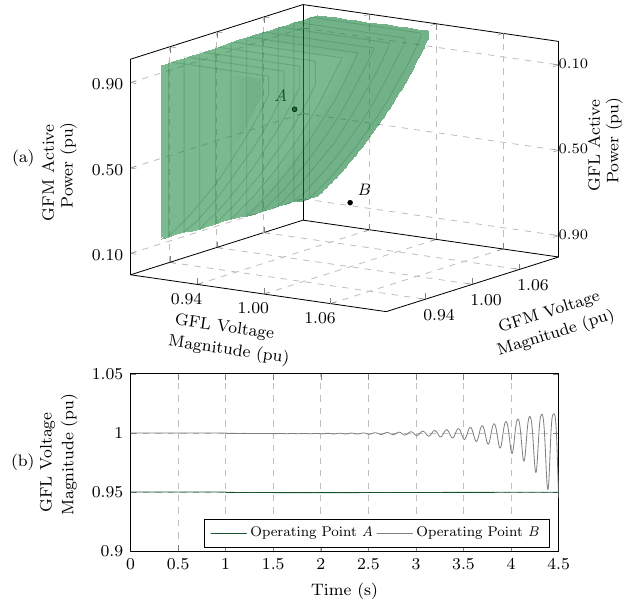}
    \caption{(a) Shaded region of stability with respect to the GFL and GFM dispatch decisions in the IEEE 39-bus system. (b) Time domain simulation for operating points a and b for a load disturbance at $t=1.00\,\rm{s}$.}
    \label{fig:39Bus_result}
    \vspace{-0.6cm}
\end{figure}

In this example, we analyze the small-signal stability of the system as we vary the active power injection of the GFM converter at Bus~36 as $p_{36}\in[0.01\,\rm{pu},1.00\,\rm{pu}]$ (of its nominal capacity) while the GFL converter at Bus~30, on the other end of the network, is varied as $p_{30}=1-p_{36}$. This allows assessing whether a higher power injection from the GFM or GFL converter is more favorable for supplying the same load.  Furthermore, the voltage magnitudes of these converter units are independently varied as $v_{36}\in[0.90\,\rm{pu},\,1.10\,\rm{pu}]$ and $v_{30}\in[0.90\,\rm{pu},\,1.10\,\rm{pu}]$. The result of this analysis is shown in Fig.~\ref{fig:39Bus_result}. From the shaded stability region, we conclude that it is generally preferred to meet the consumption by a larger share of active power from the GFM unit compared to that of the GFL unit. Moreover, the stable region is skewed towards the lower-voltage of the GFL converter. In particular, reducing the GFL voltage $v_{30}$ below $0.95\,\rm{pu}$ enlarges the admissible power injection of the GFL converter and increasing it beyond $0.95\rm{pu}$ rapidly diminishes the stable power injection. Increasing the voltage magnitude of the GFM converter increases the admissible power injection of the GFL converter, albeit marginally. The lower admissible power injections of the GFL converter with higher voltage magnitudes are well-aligned with the results obtained in the previous three-bus example.
Similar to the three-bus examples, we consider a stable operating Point~\emph{A} and an unstable operating Point~\emph{B} and perform a time-domain simulation with an active and reactive power disturbance of $0.10\,\rm{pu}$ each at $t=1.00\,\rm{s}$. The oscillatory instability observed outside the stable region, with a high participation of the dynamic states of the PLL, is exemplified in the response of the operating Point~\emph{B} in Fig.~\ref{fig:39Bus_result}(b).    

% \begin{figure}[]          
% 	\centering
% 	\begin{minipage}{0.5\textwidth}
% 		\centering
%  		\hspace{0.001cm}
% 		\scalebox{0.925}{\includegraphics[]{images/SectionIII/39bus_region.pdf}}\\
% 		\vspace{0.01cm}  
% 	\end{minipage}
% 	\caption{\label{fig:3bus_SG_Gfl_res}Region of stability with respect to active powers and voltage magnitudes of the GFM and GFL unit at Bus 31 and Bus 37, respectively. }
% 	\vspace{-0.35cm}
% \end{figure}

\section{Conclusion}\label{sec:conclusion}

In this paper, we implement and simulate detailed models of modern power system components to highlight that small-signal stability is equally affected by dispatch decisions and control parameters. We map the stable operating regions considering the eigenvalues and reveal how the magnitude of voltage, the active power, and the reactive power can influence the stability. Across test systems, we find that high voltage and capacitive reactive power GFL operation requires a reduction in its active power injection, whereas inductive operation enables higher injection of active power. In general, a higher active power injection from the GFM converter is preferred to that from the GFL converter. Although the stable operating regions obtained and their insights could be used to directly constrain the respective dispatch variables, future work will formalize optimal power flow formulations that incorporate these stable regions for converter-dominated systems.

% References section
\bibliographystyle{IEEEtran}
\bibliography{bibliography}

% That's all folks
\end{document}